# Electron Pulse Compression with a Practical Reflectron Design for Ultrafast Electron Diffraction

Yihua Wang and Nuh Gedik

*Abstract*—Ultrafast electron diffraction (UED) is a powerful method for studying time-resolved structural changes. Currently, space charge induced temporal broadening prevents obtaining high brightness electron pulses with sub-100 fs durations limiting the range of phenomena that can be studied with this technique. We review the state of the the art of UED in this respect and propose a practical design for reflectron based pulse compression which utilizes only electro-static optics and has a tunable temporal focal point. Our simulation shows that this scheme is capable of compressing an electron pulse containing 100,000 electrons with 60:1 temporal compression ratio.

*Index Terms* — electron pulse compression, reflectron, ultrafast electron diffraction, structural dynamics.

## I. INTRODUCTION TO ULTRAFAST ELECTRON DIFFRACTION

Being able to directly observe structural dynamics in a time resolved fashion is crucial to understand many phenomena across different disciplines [1]. In condensed matter physics, strong coupling between lattice degree of freedom and charge or spin degrees of freedoms leads to fascinating yet poorly understood behaviors such as simultaneous structural and electronic phase transitions etc... In chemistry, the ability to directly observe making and breaking of chemical bonds during chemical reactions is a key in figuring out reaction pathways. In structural biology, knowing the equilibrium static structures of proteins is not enough to understand their function, since they go through structural changes while performing their duties. This broad range of problems, therefore, necessitates a technique which has fast enough time resolution and can directly couple to the structural degree of freedom.

Timescales for these structural motions can be as fast as tens or hundreds of femtoseconds (fs) comparable to some fraction of periods of optical phonons in crystals. Current electronics are too slow to access these timescales. Pump-probe methods based on ultrafast lasers, on the other hand, have been routinely used to study electronic dynamics happening in fs timescales [2]. In these techniques, an ultrashort laser pulse is split into two portions: a strong portion (pump) and a weaker (probe) pulse. Both of these beams are focused on to the same spot on the sample and the relative time delay in between the two can be changed in fine steps by changing the optical path length difference. Pump pulse generates an excited state in the sample and the probe pulse is used to measure the resulting recovery dynamics by recording transient changes in optical constants such as reflectivity or transmission at different times after the photoexcitation. Time resolution in these experiments is usually only limited by the laser pulsewidth which can be as short as 10 fs. Although they provide the necessary time resolution, these optical techniques mostly reflect dynamics of electronic degree of freedom and can not provide direct information about structural motions. This is because of the fact that optical light does not directly couple to the lattice since it is absorbed by the electrons and its wavelength is too large compared to inter-atomic distances and sizes of unit cells in crystals.

In order to directly couple to the structural dynamics using the pump probe method, the pulse width should stay as short as possible but the wavelength of the probe should be comparable to the lattice spacing. This can be achieved either by using ultrafast X-rays or ultrafast electron pulses as the probe. Using electrons as the probe has several key advantages such as much higher cross section for scattering, better match of penetration depths with optical pump pulses [3] and being able to construct table-top experiments.

This work was supported by the U.S. Department of Energy award number DE-FG02-08ER46521, and in part by the MRSEC Program of the National Science Foundation under award number DMR - 0819762.

Y. H. Wang is with the Department of Physics, Massachusetts Institute of Technology, Cambridge, MA 02139 USA and the Department of Physics, Harvard University, Cambridge, MA 02138 USA (e-mail: yihua@mit.edu).

N. Gedik is with the Department of Physics, Massachusetts Institute of Technology, Cambridge, MA 02139 USA (phone: 1-617-253-3420, fax: 1-617-258-6883, e-mail: gedik@mit.edu).

Ultrafast electron diffraction (UED) is a form of pump-probe technique that can directly couple to structural dynamics using ultrashort electron pulses as the probe. The principle of UED is illustrated in Fig. 1. An ultrafast laser pulse is split into two; the first part of the laser pulse is directly focused on to the sample to create a non-equilibrium state. To probe the induced structural change, the second part is frequency tripled and focused on to a photocathode generating an ultrafast electron packet via photoelectric effect.

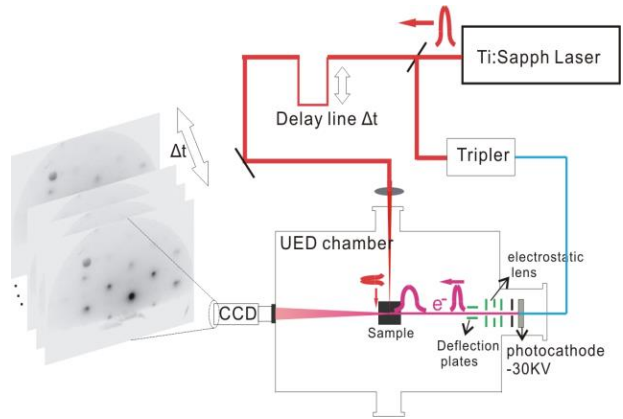

Fig. 1. Ultrafast electron diffraction setup. An amplified laser pulse (red) is split into two parts. One part goes through a delay line and into the ultra-high-vacuum chamber to excite the sample. The second part is frequency tripled and hits a photocathode. The photo-generated electrons (blue) are accelerated to high energies (typically around 30 keV) and diffracted from the sample either in reflection or in transmission geometry. An intensified CCD is used to record the diffraction pattern at different time delays.

These electrons are then accelerated through a high voltage (typically around 30 - 100 keV) and focused on to the sample. At these energies, the wavelength $\lambda$ of the electrons becomes smaller than the inter-atomic spacing (at 30 keV, $\lambda_{de\ Broglie} = 0.07$ Å) and they are diffracted from the sample. The relative arrival time of the probing electron packet and the initiating laser pulse at the sample can be changed by changing the optical path-lengths difference between the two laser beams. Recording the diffraction pattern of the electron packet as a function of this time delay provides both the equilibrium structure and a movie of the structural evolution with sub-Angstrom spatial resolution (reaching ~ 0.001 Å level) and sub-picosecond temporal resolution. UED can monitor the position, intensity and the width of the lattice Bragg spots as a function of time after the photo-excitation.

Although UED is a relatively new technique, It has already been used to successfully observe structural dynamics of many different systems in the condensed phase ranging from molecules [4] to crystalline solids and nanostructures. Some of the observed phenomena includes ultrafast non-thermal melting of simple elements [5, 6], structural dynamics in nano-structures [7, 8], bond-hardening in Au [9], photo-induced structural phase transitions in strongly correlated systems [10, 11] and measurement of electronic Grüneisen Constant [12]. Furthermore, recently we have also witnessed important developments in a closely related technique ultrafast electron microscopy (UEM) can directly record real space images of transient structures with ultrafast time resolution [13, 14].

Despite the considerable success in these experiments, there still remains a big technological challenge for UED and UEM, namely to improve the time resolution to be better than 100 fs while preserving the high brightness of electron pulses. Due to the presence of strong Coulomb repulsion in a high-density charged packet, time-width for a pulse containing only 10,000 electrons after 10 cm of travel at 30 keV is limited to several picoseconds (Fig. 2). This poses serious limitations on the study of a wide range of important physical and chemical phenomena [15] such as making-and-breaking of chemical bonds, photo-induced structural and electronic phase transitions [16] and excitation and detection of coherent optical phonons. Being able to temporally compress high density electron pulses broadened by space charge effects will open up the possibility of probing all these processes with UED.

In this contribution, we propose a practical method that allows compression of space charge broadened high brightness electron pulses down to femtosecond regime. An electron pulse freely expanding under the influence of space charge effects develops a velocity-position correlation such that the electrons at the front of the pulse speed up and the ones at the back slow down. By appropriately engineering the initial pulse profile, this correlation can be made linear and therefore reversible. Our compression method is based on the concept of a "reflectron" which electrostatically reflects the electron pulse back by having them make a U turn in space. This is done in such a way that at the output of the reflectron, faster electrons will now be at the back of the pulse and slower ones will be at the front and the two ends of the pulse will catch up right at the sample position. Our method can achieve 60:1 pulse compression and offers the advantage of being highly practical by using all electrostatic elements and by allowing tuning of the temporal focal point with a two stage reflectron design.

Below, we first discuss the nature of space charge induced broadening and briefly mention the compression schemes proposed so far. We then present analytical calculations and detailed numerical simulations of pulse compression of the proposed reflectron design.

## II. ELECTRON PULSE BROADENING AND COMPRESSION

Space-charge broadening is the biggest hurdle in achieving ultrashort electron pulses with high brightness in UED. The nature of this broadening has been studied in great detail in the literature [17-20]. To briefly illustrate this here, we performed numerical simulations tracking the free-propagation of electron pulses with various numbers of electrons in them. The photoemission process to form the initial electron pulse is simulated by randomly generating individual particles distributed with a Gaussian profile over 50 fs temporal-width and 100 μm spatial-width in the transverse direction (FWHM). We assume that initial velocities can point anywhere within the $2\pi$ solid angle. In order to calculate the two-body Coulomb interactions of N electrons, we used a numerical algorithm adopted from Ref [21] which reduces $O(N^2)$ calculation time to $O(N\log N)$. The motion of electrons is calculated by Runge-Kutta algorithm to the $5^{th}$ order [22].

The results of these simulations are shown in Fig. 2 where energy width and pulse width are plotted as a function of propagation time. The energy width experiences a very sharp rise during the first nanosecond of propagation and then levels off due to the significant decrease in electron density. The asymptotic value of the energy width increases very rapidly with the number of electrons in the pulse. The pulse width has an initial quadratic growth (Fig. 2 lower panel inset) over the first nanosecond and then increases linearly due to free expansion, signifying a much reduced space-charge effect. For a 50,000-electron pulse at 30keV, after 20cm of free propagation (2ns), the pulse width is more than 6ps, which is too long for many interesting dynamical processes [23].

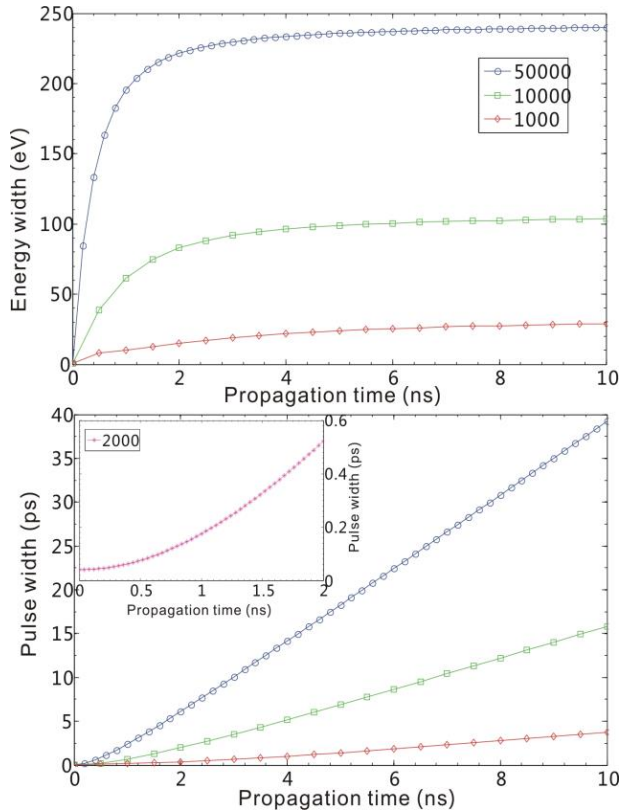

Fig. 2. Electron pulse broadening over 0-10 ns propagation time for different number of electrons (N). The initial pulse width is set to be 50 fs and radial width is 100 μm. Upper panel shows the energy broadening for N=50,000 (circle), 10,000 (square) and 1,000 (diamond). Lower panel shows the pulse-width broadening in picosecond. Inset shows the pulse-width broadening within 2 ns for 2,000 electrons.

As evident from Fig. 2, the space-charge broadening can be reduced either by decreasing the propagation time (i.e. by bringing the sample close to the photocathode) or by decreasing the number of electrons per pulse. There is a practical limit to the photocathode-sample distance in order to incorporate electron optics. If, on the other hand, one decreases the number of electrons per pulse, the repetition rate or the averaging time should be increased to collect enough electrons to form a good diffraction pattern (around $10^9$ electrons needed for this at 30 keV [5]). In order to achieve a time resolution around 100 fs with a reasonable averaging time, each pulse should contain very few electrons which necessitates the use of high repetition rate lasers (in the MHz range). This limits the energy per pulse available for the pump beam to nano-joule levels which severely limits the range of phenomena that can be studied. For example, in order to photo-induce structural phase transitions, pump fluence typically has to be more than 1 mJ/cm$^2$.

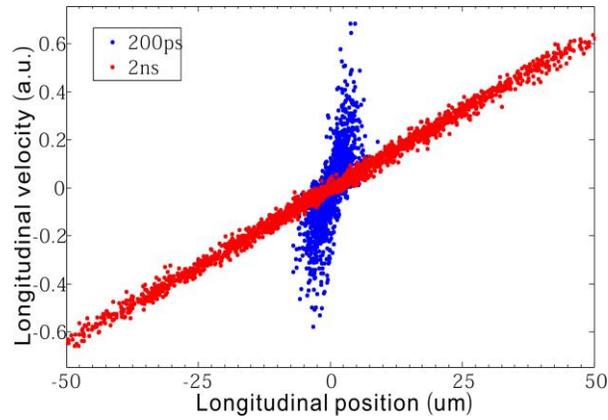

Fig. 3. Linear chirp in the longitudinal velocity vs. longitudinal position along the propogation direction of an electron pulse due to space-charge expansion. Profiles at propagation times 200 ps (blue) and 2 ns (red) are shown. The pulse depicted here is derived from a 'Top-hat' laser profile and contains 2,000 electrons. A linear chirp is well-developed after 1ns.

High brightness electron beam is also essential to UED experiments where a non-reversible ultrafast process like melting is studied. Therefore, reducing the number of electrons or decreasing the propogation time has their own limitations and temporal pulse compression is an inevitable challenge for UED. One unique feature in the phase space distribution of an electron pulse going through space-charge expansion is the possibility of forming a reversible linear chirp. As the pulse expands, the electrons at the front of the pulse speed up and the ones at the back slow down leading to a correlation of longitudinal velocity and position along the direction of propagation. It is well known that this correlation will be linear if the initial electron pulse profile is a uniform ellipsoid. (It is possible to produce uniform ellipsoid electron distribution by using a laser beam with a 'Top-hat' transverse profile rather than a Gaussian to photoinject electrons [24]). Fig. 3 shows this linear dependence between longitudinal velocity and the longitudinal position at two different propagation times (200 ps and 2 ns) calculated for a pulse containing only 2000 electrons with uniform ellipsoidal initial profile. Many different methods of compressing electron pulses that exploit this space charge induced linear chirp have been proposed. An α magnet is used to compress electron pulses [25] by letting faster electrons spend more time in an α shaped magnet. This method only works for electrons with relativistic kinetic energies and is not applicable to 30-100 keV range. Alternatively, Radio-frequency (RF) electric fields can give slower electrons at the back of the pulse a 'kick' and therefore compress the pulse [26]. It has numerically been shown that RF compression schemes are capable of achieving 30 fs pulses with a flux of 60,000 electrons and 0.5 mm spot size operating at relativistic energies. Although RF methods have great potential, they lack the technical simplicity of static electron optics and synchronization of RF cavities with laser pulses can be a big challenge. Another class of proposals utilize ultrashort laser pulses to compress electron pulses. Electrons interacting with a fast oscillating laser field experiences Kapitza-Dirac force [27]. Ultrafast lasers, due to their huge transient electric fields, can deliver a potential on the order of several eV on the length scale of their wavelength [28]. The idea of using ultrafast lasers to form a temporal potential acting as 'temporal lenses' [29] may allow compression of mildly space-charge broadened pulses (~1000 electrons) to 100 fs range or even achieve sub-fs time-width by dividing a broad pulse into bunches. It is worth mentioning that combining these compression techniques with single electron pulses may potentially extend the time resolution of UED to attosecond regime [30]. On the other hand, it would be difficult to use these techniques to generate high brightness electron pulses.

Weber et al [31] proposed using an electrostatic electron mirror called 'reflectron' to reflect and compress an electron pulses for femtosecond diffraction experiments. The idea of a reflectron is based on a simple intuition of classical physics: Under gravity it takes longer for a faster object thrown upwards to return to ground. Similarly, a faster electron entering a large enough negative potential takes longer to exit. A reflectron creates such a reflective electric potential. Faster electrons, which are ahead of slower ones at the entrance of the reflectron due to space charge effects, will be trailing behind slower ones when they exit. The pulse is self compressed as faster electrons gradually catch up in the free-fly region (see Fig. 4) in spite of the presence of space-charge repulsion. The pulse will start to broaden again once it reaches the temporal focus. If one can tune this temporal focal point to exactly match the sample position, the maximum time resolution will be obtained. Comparing with other proposals, reflectron holds the promise to produce high brightness ultrashort electron pulses in a table-top setup with non-relativistic electrons. Using only static fields, it is also easier to implement than a RF compression cavity.

In the following section, we focus on electron pulse compression using reflectron and propose a practical design.

## III. PULSE COMPRESSION WITH REFLECTRON

Historically, the technique of using an electrostatic field to reflect a charged particle bunch was first proposed and realized in mass-spectroscopy [32], where the name 'reflectron' was first used. A charged particle packet with an energy spread but the same mass can be focused in their fly time to enhance the resolution of the time-of-flight based mass spectroscopy.

Ref. [31] first proposed applying the idea of reflectron to compress electron pulses for the purpose of UED. Ref. [33] has performed detailed numerical simulations accounting for space-charge effects inside reflectron and other electron optics. Their simulation successfully showed that reflectron is able to compress both relativistic and non-relativistic pulses with high brightness (50,000 electrons and 100 μm) to sub-ps temporal width.

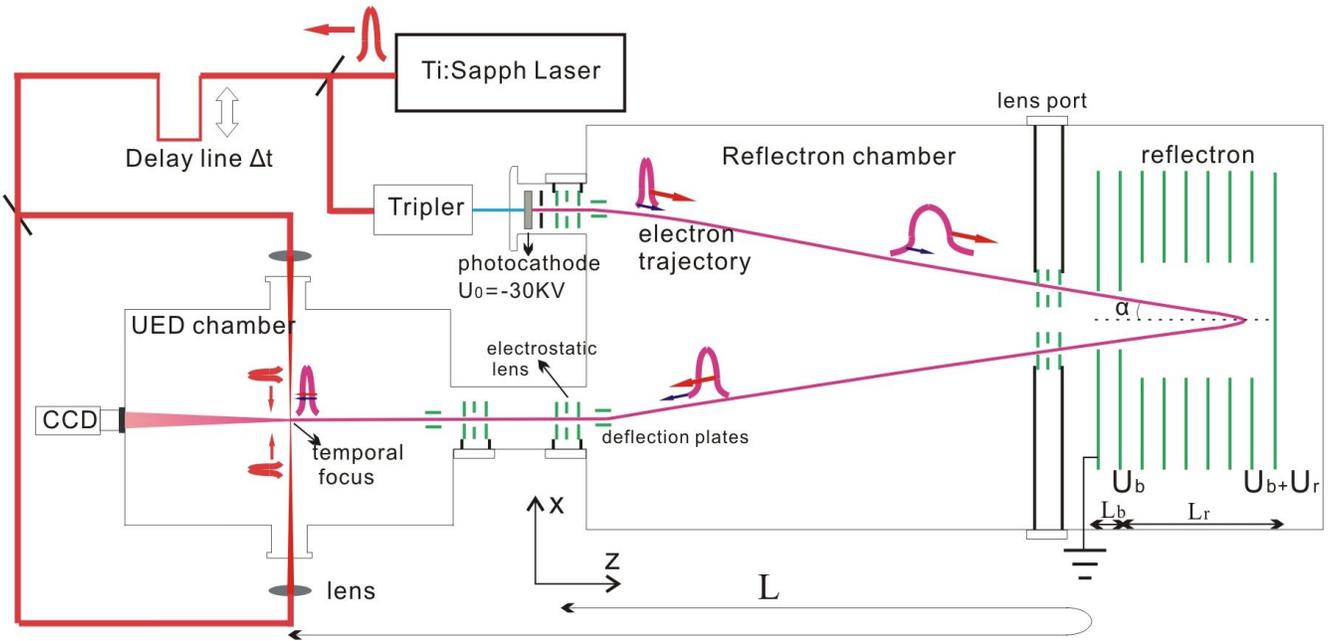

Fig. 4. Schematics of the proposed reflectron chamber connected to the UED chamber capable of measuring the compressed electron pulse-width. Electrons are generated at the photo-cathode by ultrafast UV pulses and accelerated to high energies. They are then focused by electrostatic lenses and guided by deflection plates towards the reflectron during which they experiences space-charge induced temporal broadening. After being reflected by the reflectron, the pulse starts to be compressed. More electrostatic lenses and deflection plates steer the pulse into the UED chamber and focus it at the sample position where a transient intensity grating is formed by two counter-propagating laser pulses. The Pondermotive force from the transient grating deflects the electron pulse during a cross-correlation measurement that is recorded by the CCD camera. The electron pulse compression can be quantified and optimized by changing the voltages $U_b$ and $U_r$ of the reflectron and the voltages of the lenses.

The reflectron design in [33] uses magnetic lenses and bending magnets to manipulate the pulse in conjunction with a single stage reflective electrostatic potential. Although it leads to a near perfect compression of electron pulses numerically, significant technical challenges need to be addressed to implement this design into practice. Magneto-static electron optics is difficult to incorporate into a geometry that involves a reflection trajectory. This is because solenoid coils that can focus electrons with 30 keV energy or above usually require electrical currents on the order of an ampere. As a result, they generate a huge amount of heat and need to be air or water-cooled. Since the idea of reflectron depends on a near-perfect time-reversal, a symmetric electron optics setup about the reflectron center axis is essential for achieving a compressed pulse that has a pulse-width close to its initial width at the anode. It is geometrically very difficult for magnetic lenses and benders outside the vacuum tube to be arranged in such a fashion. In-vacuum water cooling of these elements is also technically very challenging to implement with such geometry. Using bending magnets to compensate transverse velocity dispersion [33] requires precise mechanical angular alignment between the reflectron, photocathode, magnetic lenses and the bending magnets. A small misalignment will cause significant pulse broadening. Lastly, having only one reflectron stage maintaining at certain electric field only gives one degree of freedom which is used to match with the kinetic energy of the electron pulse for temporal compression. This leaves no degrees of freedoms to adjust the temporal focal length.

Here, we propose a practical two-stage reflectron design that is entirely composed of electrostatic optics to overcome the above practical limitations. Our two-stage non-magnetic design allows voltage tuning of temporal focal length, which is important considering the large increase of pulse width even slightly away from focal point. The electro-static lenses and deflection plates not only eliminate the problem of transverse dispersion but also simplify the mechanical alignment. These in combination with a recently developed scheme of in-situ pulse-width measurement [28] holds the potential to reach femtosecond time resolution in high brightness ultrafast electron diffraction.

The schematic of the experimental setup that is used for pulse compression and diffraction is shown in Fig. 4. It consists of two connected chambers (UED and reflectron chambers). Detailed trajectory of the electron pulse is described in the caption of Fig. 4.

Inside the UED chamber, a CCD is used to record the diffraction pattern and a multi-axis goniometer is used for precise rotation and translation of the sample to align with the electron beam. A transient laser intensity grating is constructed at the position of the sample by two counter-propagating ultrafast pulses. The enhanced Ponderamotive force [28] due to enhanced intensity gradient with pitch length half of the laser wavelength acts on the electron pulse transversely and enables a cross-correlation measurement of electron pulse-width. This integrated design allows in-situ optimization of the pulse-width.

The reflectron chamber functions as an evacuated 'optics table', on which different electron optics such as lenses, deflection plates and the reflectron stage can be mounted and biased through side ports. Since the space-charge broadening also affects the pulse transversely, electrostatic lenses are placed along the electron trajectory every 50 cm to regulate the pulse diameter. The distance between the lenses is chosen so that the pulse diameter does not grow bigger than the size at which the thin lens approximation is applicable. Considering that the biggest challenge in reflectron and electron optics in general is to align the electron beam through the optical axis of every component, this design also allows fine alignment of individual optic either by mechanical translational or by voltage adjustment. In our simulations, the lenses are enzel type [34] cylindrical lenses which are composed of three electrodes. The middle one is biased to a voltage comparable to the electron energy and the outer ones are grounded. The axes of the lenses need to be parallel to the axis of the reflectron so that the lens will not cause any tilting of the pulse and break the symmetry about the optical axis.

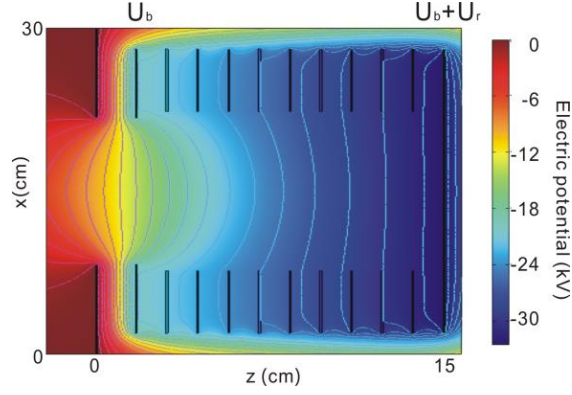

Fig. 5. Electrostatic potential of the proposed reflectron for an acceleration voltage of 30 kV. The voltage on the second ring is -12 kV and on the last ring is around -31 kV. These voltages can be adjusted to shift the temporal focal point along z direction (see text).

The deflection plates are two sets of electrostatic parallel plates controlling both x and y motions. Parallel plates sufficiently long will not tilt the pulse and can avoid broadening of longitudinal width due to tilting [32] if the photocathode and all the electrostatic lenses' axes are parallel to the axis of the reflectron. Below we will show analytically that by keeping the axes parallel, the transverse dispersion displacement is very small even for a large energy width and can safely be ignored. Our simulation of the compressed pulse also confirms that there is no transverse energy dispersion in our design. This is another advantage over using bending magnets [33] which have to be carefully balanced according to the incident angle to cancel the tilting effect.

The reflectron (Fig. 4 and 5) consists of a series of concentric rings [32]. The first one is grounded. The second and the last one are biased to voltage $U_b$ and $U_b+U_r$ (see Fig. 5). The $n$ rings in between the second and the last rings are separated by resistors so that the $i^{th}$ ring can reach a voltage between $U_b$ and $U_b+U_r$ according to the equation:

$$U_i = U_b + i \times \frac{U_r}{n+1}$$

This configuration creates two regions with different electric fields within the reflectron. The first region is between ground and $U_b$ with length $L_b$. It has a steep electric field compared with the second region which is between $U_b$ and $U_b+U_r$. It will be shown below that by tuning $U_b$ and $U_r$, the temporal focus position can easily be adjusted. We numerically solved Poisson's equation for the chosen geometries of lenses, deflection plates and reflectron to obtain their electric potential and field, which act as external force on the electron pulse. The electric potential of the reflectron is shown in Fig. 5.

The space charge repulsion is small after the initial rapid broadening as also shown in Fig. 2. Therefore, we can analytically derive the approximate voltages to set on the reflectron given the energy of the electron pulse and the chosen temporal focal point. We define the following dimensionless scaling factors of voltage and reflectron size:

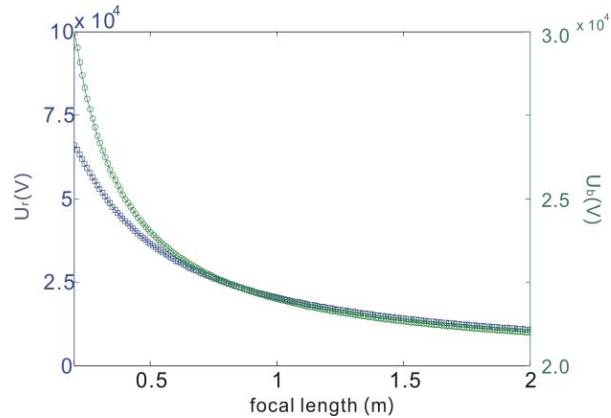

Fig. 6. Tuning temporal focal point by adjusting the reflectron voltages. The energy of the electron pulse is set to 30 keV. The lengths of reflectron stages are $L_b$ = 5cm and $L_r$ = 11cm.

$$P_b \equiv \frac{U_b}{U_0}, P_r \equiv \frac{U_r}{U_0}$$

$$Q_r \equiv \frac{L_r}{L}, \ Q_b \equiv \frac{L_b}{L}$$

, where $U_0$ is the photo-cathode potential and $L$ is the total drifting distance outside reflectron, $L_b$ and $L_r$ are the lengths of the two regions inside the reflectron respectively as indicated in Fig. 4. It can be shown that, by setting the reflectron potentials to:

$$P_b = (4Q_b + 2)/3$$

$$P_r = \frac{4Q_r P_b}{3P_b - 2[1 - (\sqrt{1 - P_b})^3]} \tag{1}$$

the travel time t of electrons drifting a distance L is independent of their energy to the second order of the relative energy width $\delta \equiv \Delta E / E$ and can be expressed as:

$$t(\delta) = 2t_0 P_b (1 + O(\delta^3))$$

where $t_0 \equiv L / \sqrt{2eU_0 / m_e}$ . Since $L_r$ and $L_b$ are fixed for given a reflectron setup, the voltage on the reflectron is adjusted according to the parameters $Q_b$ and $Q_r$ as in Eq. (1) to obtain the desired $L$. The reflectron potential has to be high enough to reflect the pulse, i.e., there is a lower limit for $Q_b$ and/or $Q_r$ so that $P_b + P_r > 1$. Regarding the reflectron as a temporal lens, $L$ is essentially the focal length for the chosen voltages. We plot out the voltage $U_b$ and $U_r$ as a function of this focal length in Fig. 6. It is clear that to focus the electron pulse further away from the reflectron, the reflectron voltages have to be smaller for a slower self-compression. In the limit of infinitely far focus, the sum of $U_b$ and $U_r$ is asymptotically close to the kinetic energy of the electron pulse.

We performed numerical simulations of the electron pulse propagation taking into account both the space-charge effects and the external electrostatic fields over the entire pulse trajectory. The sizes of the electron optics are decided according to the practical constraints of a UHV chamber and the size of the reflectron, which is about 1 m in length and 40 cm in diameter.

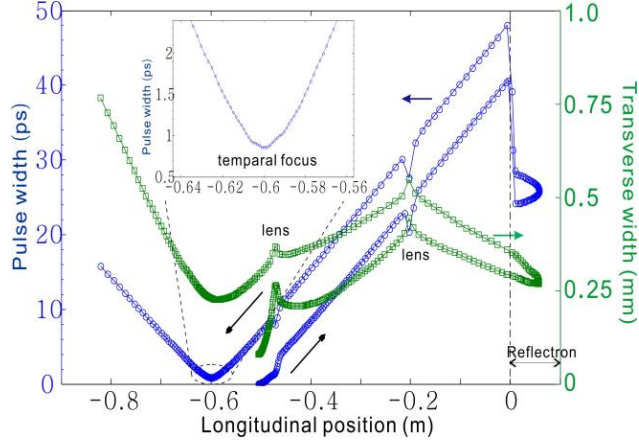

Fig. 7. Electron pulse-width (blue circles) and radial-width (green squares) as a function of longitudinal position along its trajectory in the reflectron. The pulse contains 100,000 electrons and is generated at z = -0.5 m initially moving towards +z direction. The reflectron is located between z = 0 - 0.1 m. The inset shows the magnified view of the temporal focus. The kinks on both traces come from the electrostatic lenses and the deflection plates. Note that the temporal width is obtained from longitudinal width divided by a constant speed over the entire trajectory for better visualization of the data, even though the actual temporal width is much bigger inside the reflectron due to the slow speed of the pulse. A compression ratio of 60:1 is achieved.

The two-body interactions are calculated in the same fashion as described in Sec. II. In addition, we employ an algorithm to adaptively select the integral step size depending on the electron density so that the space-charge effects can be taken into account more efficiently.

The results of these numerical simulations for 100,000 electrons are shown in Fig. 7 where the temporal width and transverse size of the electron pulse are plotted along its propagation path. Each data point corresponds to a simulation step where space-charge field is re-calculated using Barnes algorithm [21] whereas solid lines connect steps of Runge-Kutta integration [22] with much finer steps. It is clear from the higher density of data points around pulse generation and temporal compression point that the space-charge effects are properly handled.

The electron pulse is generated with 100 fs temporal width and 100 μm FWHM diameter. It travels towards positive z direction passing through the lenses and the deflection plates before entering the reflectron located in the positive z region. The effects of the electrostatic lenses are clearly visible on the kinks in the temporal width as well as in sudden reduction of the transverse size. Upon entering the reflectron, the pulse is compressed first since the fast electrons at the front of the pulse suffers a larger energy loss. The reverse process happens when the pulse exits the reflectron which results in a sudden stretching of the pulse-width. After the entire pulse moves out of the reflectron region, the pulse starts its compression towards minimal pulse width. The compression and expansion away from the high density regions show linear pulse-width versus distance dependence. At $z = -0.6\ m$ where the pulse is compressed temporally as well as transversely, the pulse-width shows a parabolic dependence on distance similar to the space-charge expansion right after the pulse is generated (Fig. 2). This is consistent with the fact that strong space-charge repulsion is present at the temporal focus. The data shows a compression ratio of 60:1 by comparing the maximal and minimal pulse-width after the reflectron stage. This high compression ratio demonstrates again the effectiveness of the two-stage design of reflectron as a temporal lens.

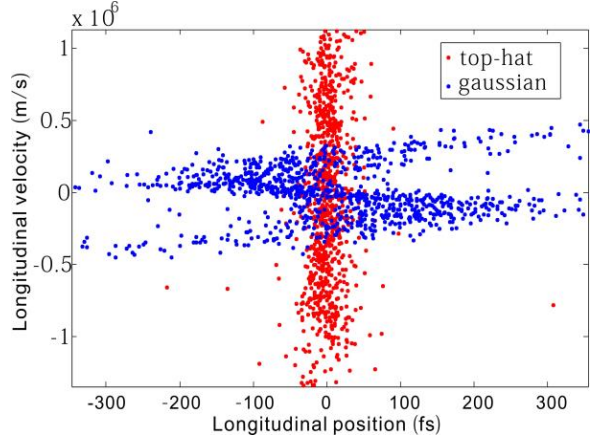

Fig. 8. Longitudinal velocity vs. longitudinal position at the temporal focus for two different electron pulses generated with a 'Top-hat' (red) and a Gaussian (blue) laser profiles. The two pulses each contain 1000 electrons and starts with 50 fs pulse-width and 100 μm radial-width (edge-to-edge for the 'Top-hat' and FWHM for the Gaussian). Longitudinal position is converted to time assuming 30 keV kinetic energy. Initial spatial charge distribution affects the achievable minimum pulse-width (50 fs for the 'Top-hat' and 400 fs for the Gaussian profile.

Our simulations show very strong coupling between the longitudinal and transverse spatial degrees of freedom at the temporal focus, especially for large number of electrons. A shorter temporal pulse means a large spot size and vice versa, indicating a strong space-charge effect at the focus. In the result shown in Fig. 7, the transverse width is kept relatively small by the electrostatic lenses at the expense of some reasonable temporal broadening to maintain a high brightness of the pulse. If the optical axes of photo-cathode, lenses, deflection plates and reflectron are all parallel, the transverse separation $\Delta c$ due to an energy difference of $\Delta E$ after reflection is:

$$\Delta c \approx L_r \tan \alpha \frac{\Delta E}{U_r}$$

The length of the reflectron tube is chosen to be 80 cm to reduce the angle α (Fig. 4) from the reflectron axis while keeping the aperture of the optical elements large enough to avoid aberration effects. Therefore, a 200 eV energy broadening (50,000 electrons) will only cause a 2 μm transverse dispersion and can be safely ignored.

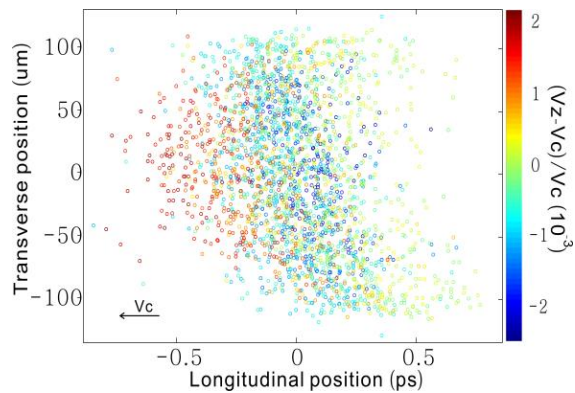

Fig. 9. Spatial velocity distribution of the compressed electron pulse at the temporal focus. The axial and radial position of 1000 randomly selected electrons from a 100,000 electron-pulse is shown. Color represents its axial velocity $V_z$ relative to its center of mass velocity $V_c$. Fast (red) electrons are a little ahead of slow (blue) electrons (pulse is propagating towards negative z direction). The slight tilting is caused by the pulse not going through the optical axis of the lenses (see text).

The principle of reflectron and most other compression schemes proposed so far relies on a well developed linear chirp. As mentioned above, this can not be achieved with a Gaussian profile and requires a uniform initial ellipsoidal distribution of electrons. [24]. To investigate the difference between these two cases, we performed reflectron compression simulations with these two initial density distributions (Fig. 8). In the case of a Gaussian distribution, the maximally compressed pulse has an 'S' shape in its phase space reminiscent of the phase space after space-charge broadening takes place [24]. This significantly limits the pulse-width. The uniform ellipsoidal distribution, on the other hand, has a well confined phase space.

Finally, we discuss potential limitations of the proposed design. Our simulations show a 500 fs compressed FWHM temporal pulse-width and 50:1 compression ratio for 50,000 electrons with 200 μm diameter FWHM. This compares with 260 fs and 10:1 compression ratio with similar parameters in Ref. [33]. We think the larger pulse-width in our design comes from the limited optimization of our numerical computation and does not indicate a fundamental limit that can be achieved with this design. The large parameter space of all the optics (geometry, position, voltage) and the long computational time to include space-charge effects makes it a computational challenge to optimize the pulse width.

The lack of global optimization causes the spherical aberration shown for 100,000 electrons in Fig. 9. The time point shown is slightly over-compressed because the fast electrons (red) are slightly ahead of the slower ones. Small tilting of the pulse is caused by the fact that the pulse does not exactly go through the center of the lenses in front of the reflectron. The most significant broadening factor is due to the spherical aberration of the lenses. The aberration is manifested in the 'bow' shape of the pulse that lacks the mirror symmetry about its vertical axis. When the electron density is high, the transverse broadening is also significant even though lenses are used to confine it. As shown in Fig. 7, the transverse width is about 0.5 mm at the lenses in front of the reflectron. The tradeoff between the size of the lenses and biasing voltage restricts the lens aperture to be around 2 mm. When the pulse enters the lens at an angle $\alpha$ about its optical axis, the parallel ray approximation is further compromised. Nevertheless, a better algorithm to perform global optimization on the positions and voltages of the electron optics can potentially overcome this problem and achieve 100 fs pulse-width for 100,000 electrons.

## IV. CONCLUSION

We presented a practical non-magnetic reflectron design in order to compress high brightness electron pulses for ultrafast electron diffraction experiments. Our design enables voltage-tuning of the temporal focal point, simplified implementation, easy alignment and in situ measurement of the compressed pulse-width at the sample position. Through detailed numerical simulations, we demonstrate that this method is capable of compressing high-brightness electron pulses down to the femtosecond regime. Practical realization of such high brightness ultrashort electron pulses will open up new avenues for UED studies of ultrafast structural dynamics in a wide range of systems.